# Prediction of Event Related Potential Speller Performance Using Resting-State EEG

Gi-Hwan Shin, Minji Lee, Hyeong-Jin Kim, and Seong-Whan Lee, *Fellow*, *IEEE*

*Abstract*— Event-related potential (ERP) speller can be utilized in device control and communication for locked-in or severely injured patients. However, problems such as inter-subject performance instability and ERP-illiteracy are still unresolved. Therefore, it is necessary to predict classification performance before performing an ERP speller in order to use it efficiently. In this study, we investigated the correlations with ERP speller performance using a resting-state before an ERP speller. In specific, we used spectral power and functional connectivity according to four brain regions and five frequency bands. As a result, the delta power in the frontal region and functional connectivity in the delta, alpha, gamma bands are significantly correlated with the ERP speller performance. Also, we predicted the ERP speller performance using EEG features in the resting-state. These findings may contribute to investigating the ERP-illiteracy and considering the appropriate alternatives for each user.

*Clinical Relevance*— The proposed predictor would be a tool to measure in advance whether speller using the brain-computer interface is suitable to use for patients with impaired behavior.

## I. INTRODUCTION

Brain-computer interface (BCI) systems enable people to communicate with others or to operate devices using neurophysiological signals in a non-muscular way [1-2]. In specific, electroencephalography (EEG) is utilized most commonly in BCIs because it is practical [3-4]. One of the EEG features in BCIs is event-related potential (ERP), which can be used to identify neural activity related to the cognitive processes [5]. This paradigm is widely utilized in many applications such as clinical diagnosis and cognitive rehabilitation [6]. However, instability of its performance across inter-subject variability remains unresolved [7]. Another problem is that some people have ERP-illiteracy, which means the use of inefficient and unsatisfactory ERP [8]. It affected a non-negligible portion of low-performance subjects and could not be operated during a long period of training. Therefore, to understand and solve the problems for practical use, it is necessary to identify and predict features related to classification performance before performing an ERP speller.

Previous studies focused mainly on the ERP speller task without considering the relationship between the resting-state EEG and the ERP speller task [6, 8]. However, during the resting-state period, the activation patterns are investigated as evidence of various network-based studies related to ERP performance. Won et al. [9] showed that the correlation between rapid serial visual presentation task features and the ERP speller area under the curve ($r = 0.53$) and the mean square error (MSE) value through linear regression (MSE = 22.35). In Li et al. [10], network properties based on coherence in resting-state EEG and P300 amplitude during ERP speller ($r = -0.60$). These previous studies proposed the predictors to reveal the underlying neural activity, but there is still difficulty in predicting features according to the brain regions and frequency bands in resting-state before performing an ERP speller.

In this study, we investigated the relationship between ERP speller performance and resting-state EEG. We hypothesized that spectral power and functional connectivity according to the four brain regions and five frequency bands in resting-state have a strong effect on ERP speller performance. Also, we predicted the ERP speller performance using linear regression. These findings indicate the possibility of as an indicator for predicting ERP speller performance. Therefore, it would help to strengthen training strategies for ERP-illiteracy subjects and to help patients for communication or cognitive rehabilitation.

## II. METHODS

### A. Experimental Procedure

The data from eight subjects (two females, average age of 23.9 ± 1.13 years) were used. All subjects had no history of any neurologic, psychiatric, and vision problems. This study was approved by the Institutional Review Board at Korea University (KUIRB-2019-0134-01) and all subjects gave written informed consent prior to the experiments.

Fig. 1 shows the experimental setup, which composed of a 4 m resting-state and ERP speller task [11]. Resting-state was performed with the eyes closed. ERP speller is designed in a 6 × 6 matrix, consisting of 30 smart home objects, 1 changes to character speller screen, and 5 different currency objects. A trial composed of 10 sequences and each sequence consists of 12 flashing stimuli. Every flash has 6 objects in random order and each object blinks twice sequentially. A maximum of 10 sequences was allotted where each set of stimuli flashed for 50 ms, followed by an inter-stimulus interval of 135 ms. To avoid repeated flashes of the same target stimulus between sequences, the presentation order was randomized and non-target stimuli were flashed consecutively.

*Research supported by Institute for Information & Communications Technology Promotion (IITP) grant funded by the Korea government (No. 2017-0-00451; Development of BCI based Brain and Cognitive Computing Technology for Recognizing User's Intention using Deep Learning).

G.-H. Shin, M. Lee, and H.-J. Kim are with the Department of Brain and Cognitive Engineering, Korea University, 145, Anam-ro, Seongbuk-gu, Seoul 02841, Republic of Korea (email: gh_shin@korea.ac.kr (G.-H. Shin), minjilee@korea.ac.kr (M. Lee), kme0115@korea.ac.kr (H.-J. Kim)).

S.-W. Lee is with the Department of Artificial Intelligence, Korea University, 145, Anam-ro, Seongbuk-gu, Seoul 02841, Republic of Korea (corresponding author: sw.lee@korea.ac.kr).

Accepted for publication at EMBC 2020. © 2020 IEEE

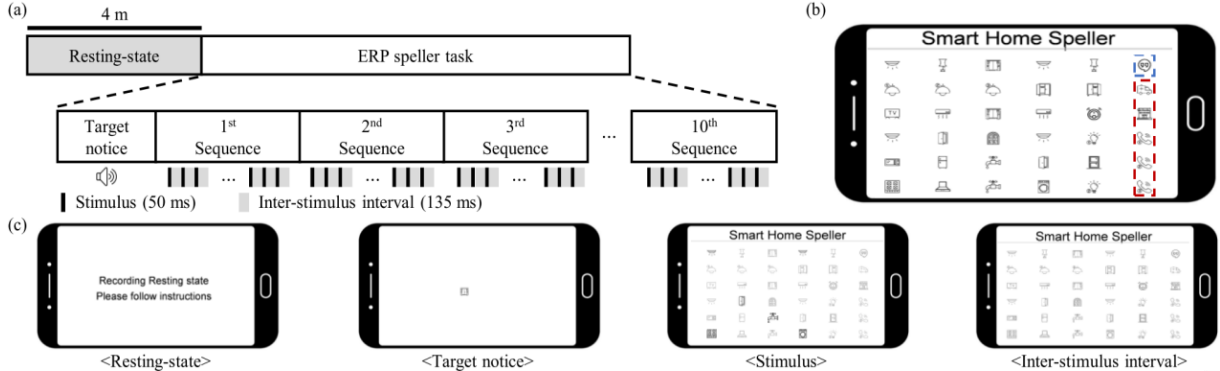

Figure 1. Experimental setup. (a) The procedure of resting-state and ERP speller task. The subjects perform 10 sequences to select a target object. (b) 6 × 6 matrices smart home speller highlighting six objects in random order. The blue dashed box changes from the smart home speller screen to the character speller screen and the red dashed box indicates different call lists. (c) It shows the smart home speller according to each condition.

## B. Data Acquisition and Preprocessing

The EEG was recorded with a sampling rate of 100 Hz and collected with 32 Ag/AgCl electrodes according to the 10-20 international system configuration. The reference electrode was placed on the nose and the ground electrode on the right mastoid. For all electrodes impedance was kept below 10 kΩ.

All data analyses were performed using MATLAB R2018b with the EEGLAB toolbox [12]. During preprocessing, EEG data were band-pass filtered between 0.5 and 45 Hz, and a notch-filter was applied 60 Hz to remove power-line interference. Then, to remove artifacts such as eye blink or muscle movement, the epochs were removed and the EEG channels were interpolated if the amplitude value (± 100 $\mu V$) was exceeded.

## C. ERP Speller Performance

The EEG was segmented from 0 to 800 ms concerning the stimulus onset and were corrected by subtracting the -200 to 0 ms pre-stimulus baseline from each trial. The mean amplitude values for all channels in each selected time window were calculated. These subject-dependent spatio-temporal feature vectors were formed with 512 dimensions (32 channels × 16 features). Then, ERP speller performance was calculated using the support vector machine classifier [13-14]. It was trained using these feature vectors in target and non-target trials. The data from the training phase was used for training the classifier and the data from the testing phase was used as a validation set.

Subjects with performance below 30% were considered as ERP-illiteracy according to previous works [15-16]. The performance was calculated from 1 to 10 sequences each for all subjects, literacy subjects, and illiteracy subjects.

## D. Prediction of ERP Performance from Resting-State

The fast Fourier transform (FFT) was performed to convert from the time into the frequency domains for spectral analysis. The frequency band was divided into delta (0.5-4 Hz), theta (4-8 Hz), alpha (8-12 Hz), beta (12-30 Hz), and gamma bands (30-45 Hz), the change in the resting-state was analyzed for each frequency band.

Power spectral density (PSD) [17] was calculated for each of the five frequencies:

$$PSD_{f_1-f_2} = 10 * \log_{10}(2 \int_{f_1}^{f_2} |\hat{x}(2\pi f)|^2 df) \qquad (1)$$

where $f_1$, $f_2$ mean lower frequency and upper frequency respectively, $\hat{x}(2\pi f)$ was obtained by FFT.

For evaluating the interaction between distinct brain regions, the phase-locking value (PLV) was calculated [18]. This is a measure commonly used to calculate functional connectivity [19]. This measure is the absolute value of the mean phase difference between the two EEG signals expressed as a complex unit-length vector as follows. Its mathematical formulation reads:

$$PLV_{i,j}(t) = \frac{1}{N} \left| \sum_{n=1}^{N} e^{-i(\varphi_i(t,n) - \varphi_j(t,n))} \right| \qquad (2)$$

where $N$ is the number of trials and $\varphi_i(t,n)$ is the instantaneous phase for EEG signal $i$ in trial $n$ at time $t$.

We grouped individual electrodes into four brain regions to analyze EEG changes between brain regions: frontal (Fp1, Fp2, F3, Fz, F4, FC1, FCz, and FC2), central (C3, C1, Cz, C2, C4, CP1, CPz, and CP2), parietal (FC5, FC6, T7, T8, CP5, CP6, P7, and P8), and occipital regions (P3, P1, Pz, P2, P4, O1, Oz, and O2). The PLV was also computed in the five frequency bands.

## E. Statistical Analysis

We performed Pearson's correlation to investigate the relationship between ERP speller performance of the highest accuracy in the sequences and the resting-state EEG. Using features with significantly correlated ERP performance in the resting-state, linear regression analysis was performed to predict ERP speller performance. The predicted features were set as the independent variables and ERP speller performance as the dependent variables. Additionally, we performed the statistical analysis of target and non-target ERP using the paired permutation test with the Bonferroni correction. The alpha level was set at 0.05 for all statistical significance.

## III. RESULTS

### A. Classification Performance

The averaged accuracy of the ERP speller according to the number of sequences is presented in Fig. 2. In the 10th sequence, the accuracy and standard error were 62.5 ± 28.1% for all subjects, 75.8 ± 15.3% for literacy subjects, and 22.5 ± 10.6% for illiteracy subjects, respectively. In conditions except for illiteracy subjects, ERP accuracy was gradually increased over the sequences.

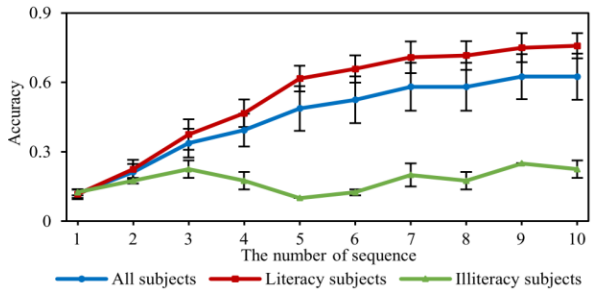

Figure 2. The mean accuracy of subjects in ERP spellers over the number of sequences. Error bars show standard error. The lines represent mean accuracy. The blue, red, and green lines indicated all subjects (n = 8), ERP-literacy subjects (n = 6), ERP-illiteracy subjects (n = 2), respectively.

### B. Correlation with ERP Speller Performance

We investigated a correlation between PSD in the resting-state EEG and ERP speller performance. Only delta PSD in the frontal region showed a strong negative correlation with ERP performance (Table I). There was no correlation between PSD in other frequency bands and classification accuracy.

Table II shows the relationships between ERP performance and PLV in five bands during resting-state EEG. In the delta band, significant correlations with performance were observed in the frontal PLV ($r = -0.857$, $p = 0.007$), frontal-central PLV ($r = -0.791$, $p = 0.019$), frontal-parietal PLV ($r = -0.846$, $p = 0.008$), frontal-occipital PLV ($r = -0.879$, $p = 0.004$), and central-parietal PLV ($r = -0.773$, $p = 0.025$). We also explored significant correlations in the frontal-central PLV in alpha band ($r = -0.726$, $p = 0.042$), frontal-central PLV in the gamma band ($r = 0.771$, $p = 0.025$). On the other hand, theta and beta bands showed no significant correlation with accuracy.

TABLE I. CORRELATION BETWEEN PSD IN RESTING-STATE AND ERP SPELLER PERFORMANCE

|  | Delta band | |
|---|---|---|
|  | *r*-value | *p*-value |
| Frontal region | -0.941 | **<0.001** |
| Central region | -0.555 | 0.153 |
| Parietal region | -0.593 | 0.122 |
| Occipital region | -0.473 | 0.237 |

### C. Prediction of ERP Speller Performance using Linear Regression

We performed linear regression using eight predictive features with a significant correlation as follows: delta PSD in the frontal region, delta PLV in frontal, frontal-central, frontal-parietal, frontal-occipital, central-parietal regions, alpha PLV in the frontal-central region, and gamma PSD in the frontal-occipital region. As a result, the MSE that represents the measure of the quality of the estimator is 22.2.

### D. Changes in Predictor during ERP Speller

Fig. 3 shows the changes in the target and non-target of the PSD and PLV with a significant correlation. The statistical differences in delta power were observed at about 200 and 500 ms in the frontal region. The statistical differences in PLV appeared between 400 to 800 ms in the frontal PLV, frontal-central PLV, frontal-occipital PLV, and central-parietal PLV in the delta band. The gamma band showed significant statistical difference at about 100 and 500 ms in the frontal-occipital PLV. On the other hand, there was no statistical difference in the frontal-parietal PLV in the delta band and the frontal-central PLV in the alpha band.

TABLE II. CORRELATION BETWEEN PLV IN RESTING-STATE AND ERP SPELLER PERFORMANCE

|  | Delta band | | Theta band | | Alpha band | | Beta band | | Gamma band | |
|---|---|---|---|---|---|---|---|---|---|---|
|  | *r*-value | *p*-value | *r*-value | *p*-value | *r*-value | *p*-value | *r*-value | *p*-value | *r*-value | *p*-value |
| F[a]-F | -0.857 | **0.007** | -0.486 | 0.222 | -0.602 | 0.114 | -0.298 | 0.473 | 0.200 | 0.636 |
| F-C[b] | -0.791 | **0.019** | -0.550 | 0.158 | -0.726 | **0.042** | -0.414 | 0.308 | 0.464 | 0.247 |
| F-P[c] | -0.846 | **0.008** | -0.086 | 0.839 | -0.364 | 0.375 | 0.074 | 0.862 | 0.621 | 0.100 |
| F-O[d] | -0.879 | **0.004** | -0.373 | 0.363 | -0.670 | 0.069 | -0.215 | 0.608 | 0.771 | **0.025** |
| C-C | -0.493 | 0.214 | -0.322 | 0.437 | -0.631 | 0.093 | 0.119 | 0.780 | 0.695 | 0.055 |
| C-P | -0.773 | **0.025** | 0.066 | 0.877 | -0.039 | 0.926 | 0.407 | 0.317 | 0.511 | 0.195 |
| C-O | -0.668 | 0.070 | -0.049 | 0.909 | -0.327 | 0.429 | 0.200 | 0.635 | 0.669 | 0.070 |
| P-P | -0.550 | 0.158 | 0.045 | 0.916 | -0.076 | 0.858 | 0.326 | 0.431 | 0.558 | 0.150 |
| P-O | -0.584 | 0.129 | -0.009 | 0.983 | -0.233 | 0.578 | 0.132 | 0.756 | 0.553 | 0.155 |
| O-O | -0.345 | 0.403 | 0.368 | 0.370 | 0.622 | 0.100 | 0.649 | 0.082 | 0.433 | 0.283 |

a. Frontal region, b. Central region, c. Parietal region, d. Occipital region.

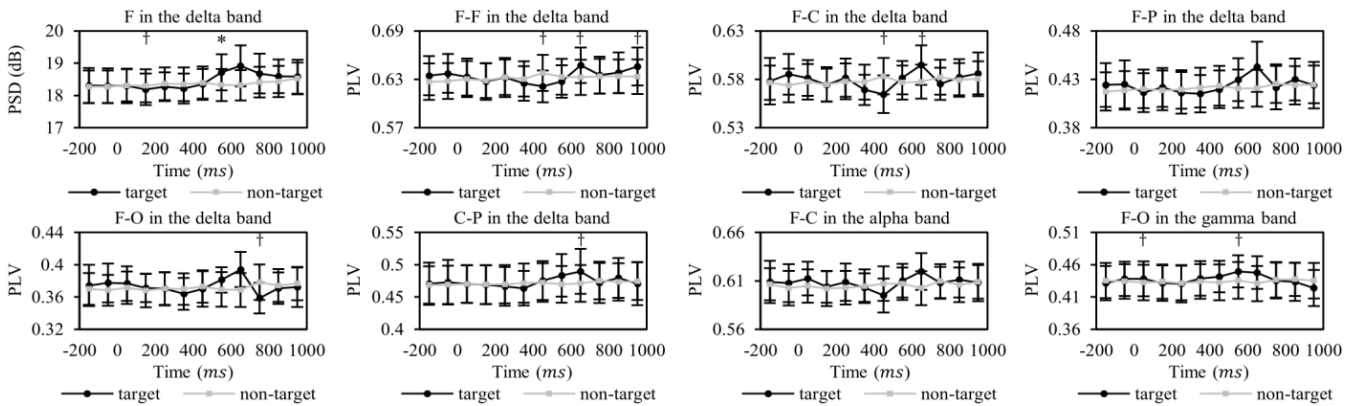

Figure 3. Target and non-target ERP distributions. The PSD and PLV indicate in each frequency band for each brain region based on the results of the significant correlation. The error bars show standard error. † < 0.05 with no correction and * < 0.05 with Bonferroni correction.

## IV. Discussion and Conclusion

In this study, we investigated the relationship between resting-state EEG and ERP speller performance. In the delta band, the spectral power and functional connectivity in the frontal region were mainly correlated with ERP performance. In addition, significantly eight predictive features were used to predict ERP performance, with large variation between subjects and the large difference in predictive features, resulting in high error value.

Brain activity in resting-state is constantly changing and it seems to affect later work [20]. Therefore, many studies have used resting-state EEG to predict BCI performance [2, 6, 9, 10, 21]. Likewise, in our study, it is thought that the cognitive function in the resting state prior to the experiment directly affected the performance.

We found that ERP performance has a negative correlation with delta power in resting-state EEG. The features of ERP in the default mode network are negatively correlated with the delta power in the frontal region [22]. It is thought that the P300 response reflects the cognitive process and this feature is dependent on the delta power associated with the cognitive role, decision making, and attentional processes [23]. Likewise, the frontal region also plays an important role in cognitive processes [24]. In this regard, PSD and PLV in the delta band before an ERP speller appears to directly affect performance.

Our study has some limitations. First, this study showed low ERP performance compared to previous studies. Obeidat et al. [25] showed low ERP accuracy using smartphones because of the small size of the stimulus and the narrow interval between each target. Therefore, our performance is not low considering the smartphone environment. Nevertheless, future studies will improve the classification accuracy for practical use. Second, the optimal EEG frequency bands may vary in different subjects and in relation to age and pathology [26], so individual frequency would be explored.

In conclusion, our study could help to better understand the inter-subject variation and ERP-illiteracy of the ERP speller using resting-state EEG. Therefore, our predictive features can be applied in a variety of ways, such as clinical diagnosis, rehabilitation, and BCIs.